\documentclass{elsart}

\usepackage{graphicx}
\usepackage{amsmath,amssymb}
\usepackage{xspace}

\newcommand{\beupl}{\ensuremath{B(\mathrm E2;~0^+_{\mathrm{gs}}\rightarrow2^+_1)}\xspace} 
\newcommand{\beup}{\ensuremath{{B(\mathrm E2)\!\!\uparrow}}\xspace}  
\newcommand{\beOneUp}{\ensuremath{{B(\mathrm E1)\!\!\uparrow}}\xspace}  
\newcommand{\ts}{\textsuperscript}

\begin{document}
\cleardoublepage
\begin{frontmatter}

%
%
\title{On the Analysis of Intermediate-Energy Coulomb Excitation Experiments}

\author[riken]{Heiko Scheit},
\ead{scheit@riken.jp}
\author[nscl,pa]{Alexandra Gade},
\author[nscl,pa]{Thomas Glasmacher},
\author[riken]{Tohru Motobayashi}

\address[riken]{RIKEN Nishina Center, RIKEN, Wako, Japan}
\address[nscl]{National Superconducting Cyclotron Laboratory, Michigan
  State University, East Lansing, Michigan, USA}
\address[pa]{Department of Physics \& Astronomy, Michigan State
  University, East Lansing, Michigan, USA}

%
%
\begin{abstract}
In a recent publication (Bertulani et al., PLB 650 (2007) 233 \cite{bertulani:2007})
the validity of analysis methods used for intermediate-energy Coulomb excitation 
experiments was called into question.
Applying a refined theory large corrections
of results in the literature seemed needed.
We show that this is not the case and that the large deviations observed 
in \cite{bertulani:2007} are due to the use of the wrong experimental parameters
in that publication.
We furthermore show that an approximate expression derived in \cite{bertulani:2007}
is in fact equivalent to the theory of Winther and Alder,
an analysis method often used in the literature.
\end{abstract}
\end{frontmatter}

\section{Introduction}
\label{intro}
Coulomb excitation (CE) of radioactive heavy-ion beams on high-$Z$ targets at intermediate to relativistic
beam energies ($E/A > 30 $ MeV) is an important and now widely used
method to study the structure of nuclei
furthest from stability and a wealth of experimental information was
obtained in laboratories worldwide
\cite{motobayashi:1995,scheit:1996,wan:1999,burger:2005,anne:1995}
since the method was first applied in 1995 \cite{motobayashi:1995}.
Even at these high beam energies it is possible to suppress the influence of
the strong interaction in the excitation process, by rejecting
events with small impact parameter, and to extract largely model
independent nuclear structure information.

Experimentally, usually the de-excitation $\gamma$ ray yields 
after CE are measured and from these the excitation cross sections are
deduced.
It is important to note that the deflection angle of the outgoing scattered
particle is restricted either by the experimental setup 
\cite{motobayashi:1995,scheit:1996,glasmacher:1997,ibbotson:1998}
or during the off-line analysis of the data, which are recorded
event-by-event \cite{gade:2003}.
The so imposed maximum scattering angle $\theta_{max}$ corresponds to a certain
minimum impact parameter $b_{min}$ which is chosen to exceed
the sum of the radii of projectile and target by
several femtometer, ensuring the dominance of the Coulomb
interaction in the excitation process.
The angle integrated CE cross section depends of course strongly on
the chosen maximum scattering angle $\theta_{max}$, which is therefore an
important parameter in the analysis.
For further details review articles \cite{glasmacher:1998,glasmacher:2001,gade:2008}
or the original references should be consulted.

In order to extract the relevant quantities, namely the reduced 
transition probabilities $B(\pi\lambda)$ and in particular \beupl values,
various analysis techniques are used:
\begin{itemize}
\item distorted-wave theory (using the ECIS code \cite{ecis-code}) used e.g. in 
  \cite{motobayashi:1995,iwasaki:2001,iwasaki:2005,chiste:2001}, 
\item the semi-classical theory of Winther and Alder \cite{winther:1979} used e.g. in \cite{scheit:1996,gade:2003,gade:2005}, and 
\item the virtual photon method \cite{bertulani:1988} used e.g. in \cite{chromik:1997}.%
  \footnote{The virtual photon method is in fact identical to the semi-classical theory of Winther and Alder.  
  Only here the excitation cross section is represented as a product of a virtual photon number and a 
  (virtual) photo absorption cross section.}
\end{itemize}

More than one method has been applied to certain nuclei and consistent
results were obtained, see 
e.g. \cite{motobayashi:1995,iwasaki:2001,pritychenko:1999,church:2005}.  
Recently, a compilation of data obtained on stable nuclides was published and a
comparison of the \beupl values obtained by various methods was given \cite{cook:2006}.
It was concluded that high-energy CE is at least as precise and accurate 
as so-called standard methods such as direct lifetime measurements, low-energy ``safe'' CE, and electron scattering.

Recently, however, a new approach was proposed \cite{bertulani:2003} and
previously published results were called into question in
\cite{bertulani:2007} claiming that
either only non-relativistic, or only partially relativistic, expressions are used or the effect of
the orbit of the nuclei, due to their Coulomb repulsion, is not fully accounted for. 
Using their new model, changes in the calculated cross sections of up to 30\% 
were reported with respect to  results from the literature.

We would like to dispute that assertion and show that the previously published results 
\cite{scheit:1996,fauerbach:1997,gade:2003} are indeed correct and consistent.

\section{Cross sections}
In \cite{bertulani:2007} Bertulani et al.\ studied the CE cross sections at
intermediate energies (10--500 MeV/$u$) using their model, published earlier in
\cite{bertulani:2003}, which aims
to correctly describe the CE process at all beam energies.
In addition, since the orbital integrals appearing in the expressions for the excitation 
amplitudes have to be determined
numerically, an approximate expression for the CE cross section was derived, called
$\sigma_{app}$ or $\sigma^{app}$ in \cite{bertulani:2007}, which is a simple
expression containing modified Bessel functions.
Using these expressions and using the numerically determined orbital integrals
it was found that the calculated excitation cross sections for a given excitation strength
are smaller by 10--30\%
in comparison to values from the literature \cite{scheit:1996,gade:2003,fauerbach:1997}
and consequently that the published \beOneUp and \beup values are in fact too small by 10--30\%.

\begin{figure}
\centerline{\includegraphics[angle=-90,width=12cm]{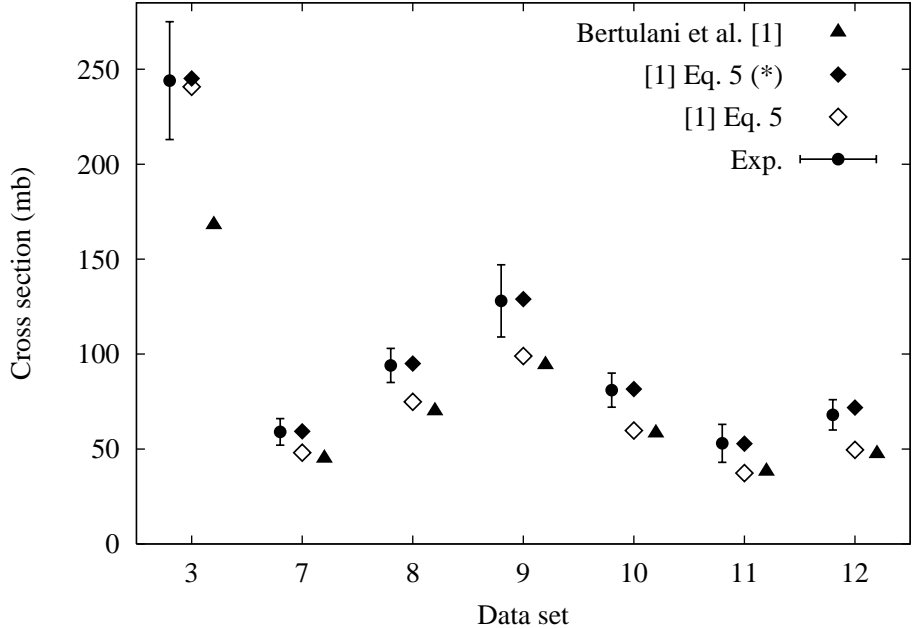}}
\caption{\label{fig}Experimental CE cross sections as listed in table \ref{tab}.
The number on the abscissa corresponds to the ``Data set'' of table \ref{tab} and
is identical to the one given in \cite{bertulani:2007}.
The circles show the experimental cross sections $\sigma_{exp}$ and the filled diamonds the cross sections
calculated by us according to Eq. (5) of \cite{bertulani:2007} ($\sigma_{app}'$ in rows with ``*'' in 
table \ref{tab}).
The open diamond shows the result of our own application of Eq. (5) of \cite{bertulani:2007} using 
the input parameters from the rows without a ``*'', i.e. the ones used in \cite{bertulani:2007}.
The triangle shows the result published in \cite{bertulani:2007}.}
\end{figure}

Since this would be a substantial alteration we tried to reproduce ($i$) our original results in 
\cite{scheit:1996,gade:2003,fauerbach:1997}
and ($ii$) the results of Ref. \cite{bertulani:2007} using Eq. (5) of \cite{bertulani:2007}.
Our findings are listed in table \ref{tab} and are displayed in figure \ref{fig}.
The number shown in column 1 of the table and plotted on the abscissa in the figure corresponds
to the ``Data set'' as given in \cite{bertulani:2007}.
Contrary to Bertulani et al.\ we essentially found exact agreement between the 
published cross sections (circles in the figure) 
and the results of our own calculation using Eq. (5) of \cite{bertulani:2007} 
(filled diamond in the figure).
However, the results reported in \cite{bertulani:2007} (triangles) deviate
significantly from those.

In order to understand the differences we performed a calculation using the 
beam energies and scattering angles
listed in table 1 of \cite{bertulani:2007}.
We would like to point out that at least for data sets number 7--12 (original publications \cite{scheit:1996,gade:2003}) 
the given values correspond to the \textit{incident} beam energy and maximum \textit{laboratory} scattering angle.
If we assume, however, that the listed beam energy is the average beam energy in the mid-plane
of the CE target%
\footnote{To determine the effective (theoretical) cross section $\sigma$ 
the beam energy dependent cross section $\sigma(E_b)$
should be averaged over the target thickness $d_t$
\[
\sigma = \frac1{d_t}\int\limits_{x=0}^{d_t} \sigma\big(E_b(x)\big)\,dx\, \approx \sigma\big(E_b(d_t/2)\big),
\]
where $E_b(x)$ is the beam energy at target depth $x$.
Due to the almost linear dependence of the cross section on beam energy within the small energy loss
in the target it is sufficient to use the beam energy in the center of the target in the calculation.}
and the angle is the scattering angle in the center-of-mass system, we can in all cases, except for \ts{11}Be,
reproduce the values given in \cite{bertulani:2007}
(see open diamonds in figure \ref{fig}).

We therefore conclude that mentioned discrepancies are due to an
interchange of center-of-mass and laboratory angles
and the use of the incident beam energy instead of the mid-target beam energy.

We furthermore found that the expressions given for $\sigma_{app}$ in \cite{bertulani:2007}
are in fact analytically identical to the one of Winther and Alder \cite{winther:1979} (see Appendix).
Since for all experimental cross sections listed in table 1 and shown in figure 1 the semi-classical theory 
of Winther and Alder was employed during the analysis of the data,
the $\sigma_{app}$ results \textit{must} be identical to the original results $\sigma_{exp}$
provided the proper input parameters are used.

This conclusion is in fact supported by an earlier work of Bertulani et al. 
\cite{bertulani:2003} where it is shown in figure 3 (curve RR) that for the case of an $E2$ excitation
the cross section determined within the Winther and Alder theory deviates by less
than 5\% from the exact treatment for beam energies above 30 MeV/$u$.

\begin{table}
\begin{tabular}{|cr|lcccccccc|}
\hline
&\# & Iso. & Ref. & $E_{beam}^{m-t}$ & $\theta_{max}^{cm}$ & $E_\gamma$ & $B(E\lambda)$ & $\sigma_{exp}$ 
                                                                                    & $\sigma_{app}$ & $\sigma_{app}^\prime$ \\
\hline\hline
  &   3 & \ts{11}Be & \cite{fauerbach:1997} & {\bf57.6}    & {\bf 3.80}  & 0.320 & 0.079 & 244 & {\bf 168.0}& 240.7 \\
* &   3 & \ts{11}Be & \cite{fauerbach:1997} & 57.6         & { 4.02}     & 0.320 & 0.079 & 244 & ---        & 245.1 \\
\hline
  &   7 & \ts{38}S  & \cite{scheit:1996}    & {\bf39.2}    & {\bf 4.10}  & 1.292 & 235   &  59 & {\bf 45.0} &  48.0 \\
* &   7 & \ts{38}S  & \cite{scheit:1996}    & {34.6}       & { 4.92}     & 1.292 & 235   &  59 & ---        &  59.3 \\
\hline
  &   8 & \ts{40}S  & \cite{scheit:1996}    & {\bf39.5}    & {\bf 4.10}  & 0.891 & 334   &  94 & {\bf 70.0} &  74.8 \\
* &   8 & \ts{40}S  & \cite{scheit:1996}    & {35.3}       & { 4.96}     & 0.891 & 334   &  94 & ---        &  95.0 \\
\hline
  &   9 & \ts{42}S  & \cite{scheit:1996}    & {\bf40.6}    & {\bf 4.10}  & 0.890 & 397   & 128 & {\bf 94.3} &  98.9 \\
* &   9 & \ts{42}S  & \cite{scheit:1996}    & {36.6}       & { 5.00}     & 0.890 & 397   & 128 & ---        & 128.9 \\
\hline
  &  10 & \ts{44}Ar & \cite{scheit:1996}    & {\bf33.5}    & {\bf 4.10}  & 1.144 & 345   &  81 & {\bf 58.3} &  59.7 \\
* &  10 & \ts{44}Ar & \cite{scheit:1996}    & {30.9}       & { 5.04}     & 1.144 & 345   &  81 & ---        &  81.6 \\
\hline
  &  11 & \ts{46}Ar & \cite{scheit:1996}    & {\bf35.2}    & {\bf 4.10}  & 1.554 & 196   &  53 & {\bf 38.2} &  37.3 \\
* &  11 & \ts{46}Ar & \cite{scheit:1996}    & {32.8}       & { 5.08}     & 1.554 & 196   &  53 & ---        &  52.8 \\
\hline
  &  12 & \ts{46}Ar & \cite{gade:2003}      & {\bf76.4}    & {\bf 2.90}  & 1.554 & 212   &  68 & {\bf 47.4} &  49.6 \\
* &  12 & \ts{46}Ar & \cite{gade:2003}      & {73.2}       & { 3.62}     & 1.554 & 212   &  68 & ---        &  71.8 \\
\hline
\end{tabular}
\caption{\label{tab}Coulomb excitation cross sections are shown in the three columns 
on the right-hand side with $\sigma_{exp}$ being the experimental CE cross section, $\sigma_{app}$ 
the cross section from reference \cite{bertulani:2007} and $\sigma_{app}'$ the cross section
calculated by us using Eq. (5) of \cite{bertulani:2007}.
The $B(E\lambda)\!\!\uparrow$ values are listed in units of $e^2$fm$^{2\lambda}$.
Values in bold were taken verbatim from \cite{bertulani:2007} and
the ``Data set'' number \# corresponds to the one used in \cite{bertulani:2007}.
In the rows starting with ``*'' the correct center-of-mass scattering angles  $\theta_{max}^{cm}$
and mid-target (m-t) beam energies $E_{beam}^{m-t}$ are listed; an agreement of $\sigma_{exp}$ and $\sigma_{app}'$ is
evident for all cases.
The rows not starting with ``*'' list the values as given in \cite{bertulani:2007} and
$\sigma_{app}'$ is calculated by us using these values.
In most cases $\sigma_{app}$ and $\sigma_{app}'$ agree, but
$\sigma_{exp}$ and $\sigma_{app}$ do not.}
\end{table}

We would also like to note that the analysis of CE data using a simple maximum scattering
angle is not applicable for all experiments in the literature.
For instance for the analysis of the CE of \ts{32}Mg \cite{motobayashi:1995}
the use of a sharp cutoff $\theta_{max}$ is meaningless (``Data set'' 6 in \cite{bertulani:2007}). 
Here the particle detection efficiency depended (smoothly) on the scattering angle
with maximum values of about 80\%.  
In this case the differential cross section must first be multiplied by the experimental 
particle detection efficiency before integrating over the scattering angle.  
This can only be done when the efficiency curve is known, which 
is shown in figure 2 (lower panel) in \cite{motobayashi:1995}.

We furthermore found the following errors in \cite{bertulani:2007}:
\begin{itemize}
\item There is a misprint in Eq. (5).  
For the E2 excitation cross section the power of $E_\gamma$ should be 2 ($E_\gamma^2$) and not 3 ($E_\gamma^3$).
\item The wrong citation was used for \ts{38,40,42}S and \ts{44,46}Ar.  
It should be Scheit et al. \cite{scheit:1996} and not Chromik et al. \cite{chromik:1997}.  
\item 
An M1 excitation was considered for \ts{17}Ne, but the excitation process proceeds
through an E2 transition with only a very weak M1 component.  
The decay, though, is an almost pure M1 transition.
\item The deviation of the cross sections published in \cite{bertulani:2007} (triangle in figure
\ref{fig})
for \ts{11}Be (``Data set'' 3) seems to be due to a forgotten factor 
of $e^2 = \alpha\hbar c \approx 197/137 \approx 1.44$ in \cite{bertulani:2007}.
\end{itemize}

\section{Conclusion}
We conclude that the theory presented in \cite{bertulani:2007,bertulani:2003} does not
represent a significant improvement over analysis methods used so far.
In fact, a good approximation \cite{bertulani:2007} to the full theory is equivalent to the theory of
Winther and Alder \cite{winther:1979}, which is widely used in the analysis of
CE data at high beam energies.
We furthermore conclude that the large differences 
of up to 30\% reported in \cite{bertulani:2007} are likely solely due to an interchange of
center-of-mass and laboratory angles and the use of incident beam energy instead of the
mid-target beam energy in the calculation of the CE cross sections.

We nevertheless agree with the authors of \cite{bertulani:2007} that
a refined theory which includes relativistic effects and
properly treats the orbit of the projectile is needed and should be applied in the future,
when high-precision results become available at next-generation radioactive
nuclear beam facilities.
The currently available data will be hardly affected due to rather large statistical and
systematical uncertainties.

\section{Acknowledgement}
This work was supported by the US National Science Foundation under grant PHY-0606007.

\appendix
\section{Coulomb excitation cross sections}\label{sec:app}
In the following $\sigma_B$ denotes the expression for $\sigma_{app}$ from
Bertulani et al. \cite[Eq. 5]{bertulani:2007}
and $\sigma_{W/A}$ the result given by Winther and Alder \cite{winther:1979}.
For a given multipolariy $\lambda$ these two expressions for the CE cross section 
differ explicitly only  in the expressions containing the modified Bessel functions $K_n$ and we will consider 
the ratio of the two expressions leaving out explicit common pre-factors.
In particular for the E2 case:
\begin{align}\label{eq:comp}
\frac{\sigma_B^{(E2)}}{\sigma_{W/A}^{(E2)}} &= \frac{\frac{8\pi^2}{75} \beta^{-4}
                       \left(2\gamma^{-2} K_1^2 + \xi\Big( 1+\gamma^{-2}\Big)^2 K_0K_1 - 
                       \frac{\beta^4\xi^2}2 (K_1^2 - K_0^2)\right)}{
  \sum\limits_{\mu=-2,\ldots2} \left|G_{E2\mu}(\beta^{-1})\right|^2 g_\mu(\xi)}\,,
\end{align}
where the functions $g_\mu$ and $G_{\pi\lambda\mu}$ are given in \cite[Eqs. 2.16 and 3.4]{winther:1979}.
The former contain the Bessel functions $K_{\mu,\mu+1}$ and the latter is a polynomial function of $\beta^{-1}$.
After a lengthy, but straight-forward, calculation one obtains:
\begin{align}
\newcommand{\ig}{\gamma^{-2}}
\begin{split}
\sum\limits_\mu \left|G_{E2\mu}(\beta^{-1})\right|^2 g_\mu(\xi) = \frac{4\pi^2\xi}{75\beta^4}
  \Big(
     -3\xi\ig K_0^2 - \xi(\beta^4+\ig)K_1^2 + \\\xi(\beta^4 + 3\ig)K_2^2 - (8\ig + 2\beta^4)K_1K_2 +
     \xi\ig K_3^2 - 4\ig K_2K_3
  \Big)\,.
\end{split}
\end{align}
Using the recurrence relations for the Bessel functions $K_n$ \cite{abramowitz-stegun}
\begin{align}
K_0 - K_2 &= -\frac2\xi K_1\qquad\text{and} & -K_1+K_3 &= \frac4\xi K_2
\end{align}
the Bessel functions $K_\mu$ with $\mu>1$ can be expressed in terms of $K_0$ and $K_1$
\begin{align}
K_2 &= \frac2\xi K_1 + K_0\\
K_3 &= \frac4\xi K_2 + K_1 = \left(\frac8{\xi^2} + 1\right) K_1 + \frac4\xi K_0.
\end{align}
The result is
\begin{align}\label{eq:wa-final}
\newcommand{\ig}{\gamma^{-2}}
\begin{split}
\sum\limits_\mu \left|G_{E2\mu}(\beta^{-1})\right|^2 g_\mu(\xi) = \frac{4\pi^2}{75\beta^4}
  \Big(\beta^4\xi^2 K_0^2 + (-\xi^2\beta^4 + 4 \ig) K_1^2 + 2\xi(1+\ig)^2K_0K_1
  \Big)\,,
\end{split}
\end{align}
where $\beta^4 + 4\gamma^{-2} = (1+\gamma^{-2})^2$ was used.
The last result (\ref{eq:wa-final}) is identical to the numerator of (\ref{eq:comp}) and
therefore
\begin{align}
\sigma_B^{(E2)} &= \sigma_{W/A}^{(E2)}\,.
\end{align}

Similarly for the E1 case:
\begin{align}\label{eq:comp-e1}
\frac{\sigma_B^{(E1)}}{\sigma_{W/A}^{(E1)}} &= \frac{\frac{32\pi^2}{9} \beta^{-2}
                       \left(\xi K_0 K_1 - \frac{\beta^2\xi^2}{2} (K_1^2 - K_0^2)\right)}{
  \sum\limits_{\mu=-1,0,1} \left|G_{E1\mu}(\beta^{-1})\right|^2 g_\mu(\xi)}\,,
\end{align}
and
\begin{align}
\newcommand{\ig}{\gamma^{-2}}
\begin{split}
\sum\limits_\mu \left|G_{E1\mu}(\beta^{-1})\right|^2 g_\mu(\xi) = \frac{16\pi^2\xi^2}{9} 
  \Big(
     (\beta^{-2}-1)(K_1^2 - K_0^2) + \beta^{-2}(-K_1^2 + K_2^2 - \frac2\xi K_1K_2)
  \Big)\,.
\end{split}
\end{align}
Using the above recurrence relations for the Bessel functions we obtain
\begin{align}\label{eq:wa-final-e1}
\begin{split}
\sum\limits_\mu \left|G_{E1\mu}(\beta^{-1})\right|^2 g_\mu(\xi) = \frac{16\pi^2\xi^2}{9}
  \Big( K_0^2 -  K_1^2 + \frac2\xi\beta^{-2}K_0K_1
  \Big)\,.
\end{split}
\end{align}
The last result (\ref{eq:wa-final-e1}) is identical to the numerator of (\ref{eq:comp-e1}) and
therefore
\begin{align}
\sigma_B^{(E1)} &= \sigma_{W/A}^{(E1)}
\end{align}
also for the E1 case.

\bibliographystyle{elsart-num}
\bibliography{cb}

\end{document}